\documentclass[mathleft
]{an}
\usepackage{graphicx}
\usepackage{times}
\def\simgt{\lower 2pt \hbox{$\, \buildrel {\scriptstyle >}\over {\scriptstyle \sim}\,$}}
\def\simlt{\lower 2pt \hbox{$\, \buildrel {\scriptstyle <}\over {\scriptstyle \sim}\,$}}
\def\athena{{\it Athena\/}}
\def\chandra{{\it Chandra\/}}
\def\erosita{{\it eROSITA\/}}
\def\euclid{{\it Euclid\/}}

\def\jwst{{\it {\it JWST}\/}}
\def\rosat{{\it ROSAT\/}}

\def\wfirst{{\it WFIRST\/}}
\def\xmm{{\it XMM-Newton\/}}
\def\xrs{{\it X-ray Surveyor\/}}

\def\xray{{\hbox{X-ray}}}

\def\aox{$\alpha_{\rm ox}$}
\def\daox{$\Delta\alpha_{\rm ox}$}

\begin{document}

\Pagespan{1}{}
\Yearpublication{2016}%
\Yearsubmission{2016}%
\Month{09}%
\Volume{1}%
\Issue{1}%

\title{High-redshift AGNs and the next decade of Chandra and XMM-Newton}

\author{W.N. Brandt\inst{1,2,3}\fnmsep\thanks{Corresponding author:
  \email{wnbrandt@gmail.com}\newline}
\and  F. Vito\inst{1,2}
}
\titlerunning{High-redshift AGNs and the next decade of Chandra and XMM-Newton}
\authorrunning{W.N. Brandt \& F. Vito}
\institute{
Department of Astronomy and Astrophysics, 525 Davey Lab,
The Pennsylvania State University, University Park, PA 16802, USA
\and
Institute for Gravitation and the Cosmos, The Pennsylvania
State University, University Park, PA 16802, USA
\and
Department  of  Physics,  104  Davey  Laboratory,  The  Pennsylvania 
State University, University Park, PA 16802, USA}

\received{23 September 2016}
\accepted{23 September 2016}
\publonline{later}


\abstract{%
We briefly review how \xray\ observations of high-redshift
active galactic nuclei (AGNs) at \hbox{$z=4$--7} have played a critical 
role in understanding their basic demographics as well as their physical 
processes; e.g., absorption by nuclear material and winds, accretion rates, 
and jet emission. We point out some key remaining areas of uncertainty, 
highlighting where further \chandra\ and \xmm\ observations/analyses, 
combined with new multiwavelength survey data, can advance understanding
over the next decade.}

\maketitle

\section{Introduction}

Over the past $\approx 17$~yr, the observational capabilities of \chandra\ 
and \xmm\ have allowed a large expansion, by more than an order of magnitude, 
in the number of \xray\ detected active galactic nuclei (AGNs) at \hbox{$z=4$--7}. 
This has come about via two primary routes. First, these missions have obtained
follow-up observations in the \xray\ regime of high-redshift 
AGNs previously found in other multiwavelength surveys
[e.g., the Sloan Digital Sky Survey (SDSS), the Palomar Sky Survey (PSS), and 
the Faint Images of the Radio Sky at Twenty-Centimeters (FIRST) survey]. 
Second, these missions have discovered new \xray\ selected high-redshift
AGNs in their multiple \xray\ surveys. 
According to a regularly updated public list compiled by 
Brandt \& Vignali\footnote{http://www2.astro.psu.edu/users/niel/papers/highz-xray-detected.txt}, 
there are now 153 \xray\ detections of AGNs at \hbox{$z=4$--7}, allowing 
reliable basic \xray\ population studies into the reionization era. Most \xray\ 
detections at \hbox{$z=4$--7} have come from the first route described above. 
However, new \xray\ discoveries from the second route continue to advance 
rapidly, and these are extremely important because they mitigate many of the
selection biases (e.g., due to obscuration and host-galaxy dilution) arising 
from optical/UV AGN selection.

In this paper, we will briefly review some of the insights that \xray\ studies
have provided about the first growing supermassive black holes (SMBHs) 
in the Universe. In \S2 we will discuss \xray\ surveys and AGN demographics, 
and then in \S3 we will cover \xray\ spectroscopy and AGN physics. For
each of these topics, we will highlight areas of uncertainty and how these
could be addressed in the next decade with \chandra\ and \xmm\  
observations/analyses.

Owing to space limitations, complex details will often need to be 
suppressed and citations cannot be complete but just representative. 
Please check the cited papers and relevant recent reviews 
(e.g., Brandt \& Alexander 2015; Reines \& Comastri 2016) for 
further references.

\section{X-ray surveys and high-redshift AGN demographics}

\subsection{Current results}

\begin{figure}
\includegraphics[width=80mm,height=70mm,angle=0]{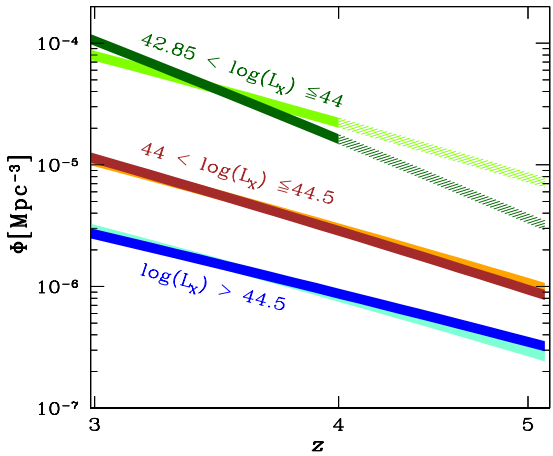}
\caption{X-ray selected AGN space density in three color-coded bins
of \hbox{2--10~keV} luminosity, derived using a selection of \xray\ surveys
that sample the luminosity-redshift plane. The lighter colors are for a pure density
evolution model, and the darker colors are for a luminosity dependent density 
evolution model. The gridded parts of the green stripes indicate a 
region where the space density is highly uncertain. Note the decline
in space density with redshift for all three luminosity bins.  
Adapted from Vito et~al.\ (2014), where the correction for redshift
incompleteness is described.}
\label{spacedensities}
\end{figure}

In contrast to early suggestions from \rosat\ surveys, current surveys
with \chandra\ and \xmm\ clearly find an exponential decline in the
space density of luminous 
\hbox{($L_{\rm X}\simgt 10^{44}$~erg~s$^{-1}$)}
AGNs at $z>3$ (e.g., see Fig.~\ref{spacedensities}). 
The decline is often modeled as $\Phi(z)\propto (1+z)^p$~Mpc$^{-3}$ with 
$p\approx -6.0$. Space-density comparisons between luminous \xray\ selected 
AGNs and optically selected quasars indicate agreement to within factors
of \hbox{$\approx 2$--3} (e.g., McGreer et~al.\ 2013; Marchesi et~al.\ 2016). 

At lower luminosities of 
\hbox{$L_{\rm X}\approx 10^{43}$--$10^{44}$~erg~s$^{-1}$}
(for the \hbox{2--10~keV} band), the space-density evolution 
is quantitatively less clear at high redshifts owing to a number of 
factors: small sample sizes, follow-up and completeness challenges, and 
sensitivity of the results to analysis details. However, at least qualitatively
all the latest results support a decline in space density at $z>3$ also for
these objects. For example, Vito et~al.\ (2014) find that the rate of decline in 
space density is plausibly consistent with that at higher luminosities, while
Georgakakis et~al.\ (2015) suggest the decline rate from \hbox{$z=3$--4}
to \hbox{$z=4$--5} may be even faster at low luminosities than at high luminosities. 

\begin{figure}
\begin{center}
\includegraphics[width=70mm,height=50mm,angle=0]{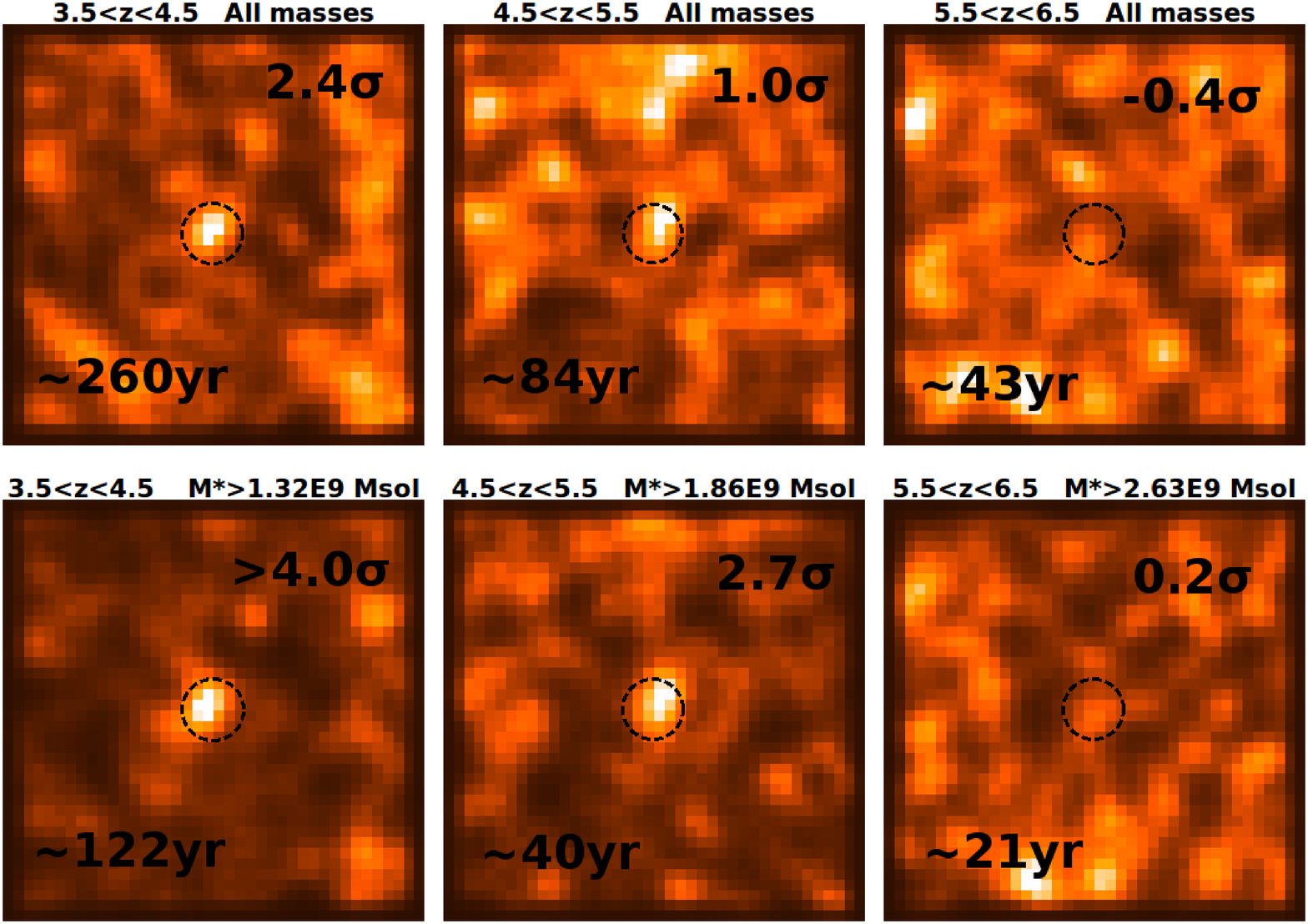}
\includegraphics[width=80mm,height=70mm,angle=0]{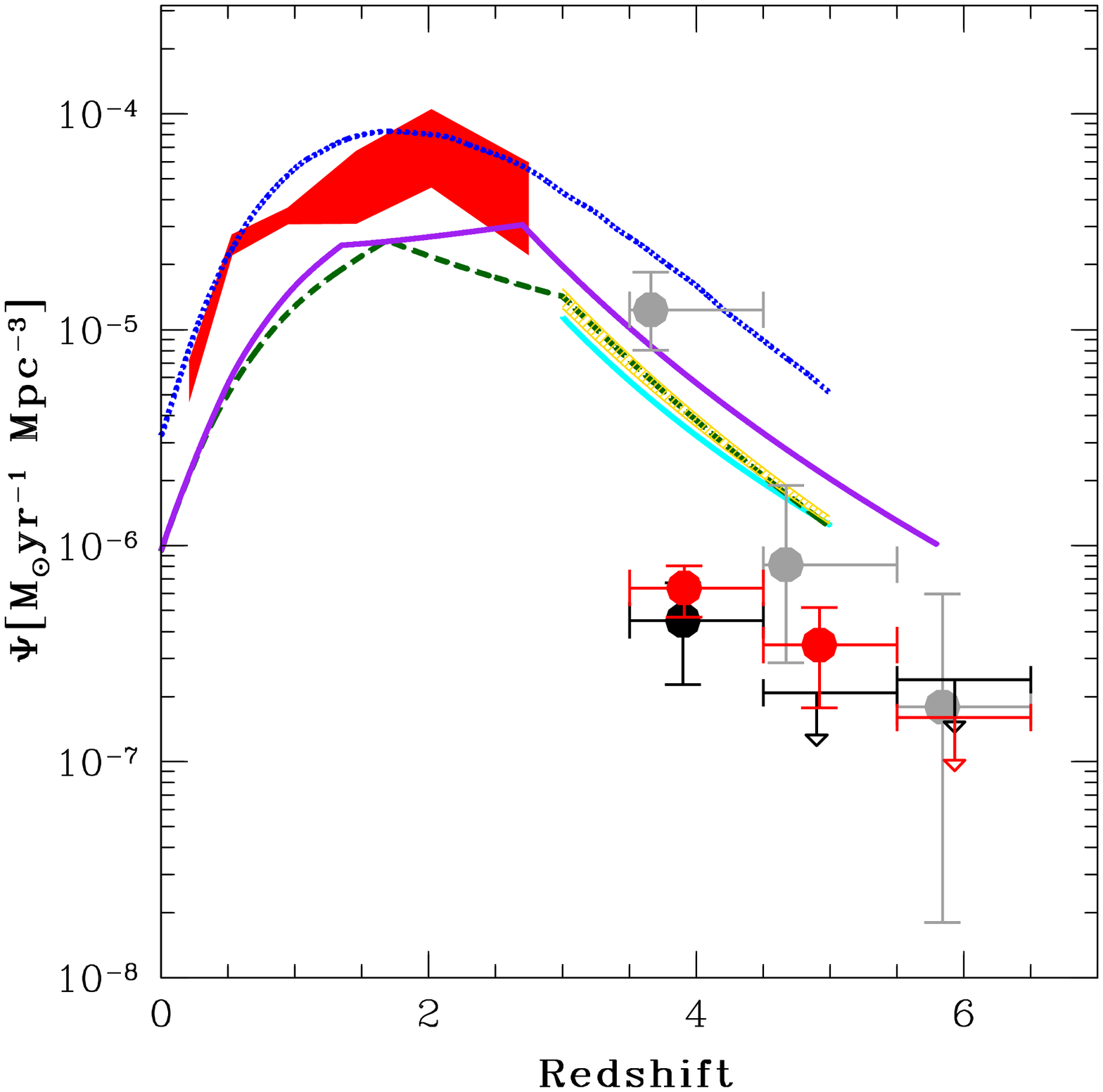}
\end{center}
\caption{{\it (Top)} Stacked \hbox{0.5--2~keV} \chandra\ \hbox{CDF-S}
images for samples of (individually \xray\ undetected) high-redshift 
galaxies. Dashed circles are centered at the positions of the stacked
galaxies with radii corresponding to the median extraction radius used
for the relevant sources. The upper panels show results for all available galaxies in 
the listed redshift ranges, while the lower panels show only the most-massive 
half of the galaxy samples. Stacked \xray\ detections are obtained 
at \hbox{$z=3.5$--4.5} and \hbox{$z=4.5$--5.5} (the detection significance 
levels are listed), while tight upper limits are obtained for 
\hbox{$z=5.5$--6.5} and higher redshifts. Each panel also lists the 
corresponding total stacked exposure time, ranging from \hbox{21--260~yr}. 
Adapted from Vito et~al.\ (2016). 
{\it (Bottom)} Constraints on the SMBH accretion-rate density vs.\ redshift derived
from the stacking results for all galaxies (black points and upper limits)
and for the most massive half (red points and upper limits). Gray points
show the (dominant) contribution from \xray\ detected AGNs in the CDF-S. The 
findings are compared with a selection of \hbox{$z=0$--5} observational results. 
The data demonstrate that most high-redshift SMBH growth occurs in the
short AGN phase; any continuous low-rate accretion contribution appears
small. Adapted from Vito et~al.\ (2016), where many further details are provided.}
\label{stacking}
\end{figure}

It is also critical to constrain the amount of high-redshift SMBH growth that may 
be occurring at still lower luminosities, corresponding to fluxes below the detection 
limits of even the deepest current \xray\ surveys. Such SMBH growth could occur, 
e.g., in a continuous or near-continuous low-rate mode, in contrast to the shorter
high-rate mode associated with individually \xray\ detected AGNs. This type of 
SMBH growth can be usefully constrained with \xray\ stacking studies, where
the \xray\ emission from hundreds of individually undetected galaxies is 
co-added. Fig.~\ref{stacking} shows some current results derived from the
deepest \xray\ survey to date, the 7~Ms \chandra\ exposure of the \chandra\ 
Deep Field-South (CDF-S; Luo et~al.\ 2017), where stacked detections are achieved up to 
\hbox{$z=4.5$--5.5} and tight upper limits are set at still higher redshifts. 
Each galaxy sample stacked in Fig.~\ref{stacking} contains \hbox{100--1300} 
galaxies, and impressively large stacked exposure times of \hbox{21--260~yr} 
are reached. The detected \xray\ signals are plausibly consistent with 
expectations for high-redshift \xray\ binary populations, according to the 
latest constraints on these populations at lower redshifts by 
Fragos et~al.\ (2013) and Lehmer et~al.\ (2016). Thus, there
is no detected evidence for SMBH growth in a continuous low-rate mode, 
and the quantitative constraints indicate this mode is not
significant at least out to \hbox{$z\approx 6$}
(see the bottom panel of Fig.~\ref{stacking}).

Collectively, the combined demographic constraints from individual \xray\ detections
and stacking indicate that AGNs are unlikely to drive reionization at $z\approx 6$, 
leaving stars as the most-likely culprit (e.g., Georgakakis et~al.\ 2015; Vito et~al.\ 2016). 
AGNs may, however, play a secondary role in distributed heating of the intergalactic
medium (e.g., Grissom et~al.\ 2014). 

\subsection{Massive X-ray archive mining enabled by new very wide field surveys}

\begin{figure}
\includegraphics[width=70mm,height=82mm,angle=-90]{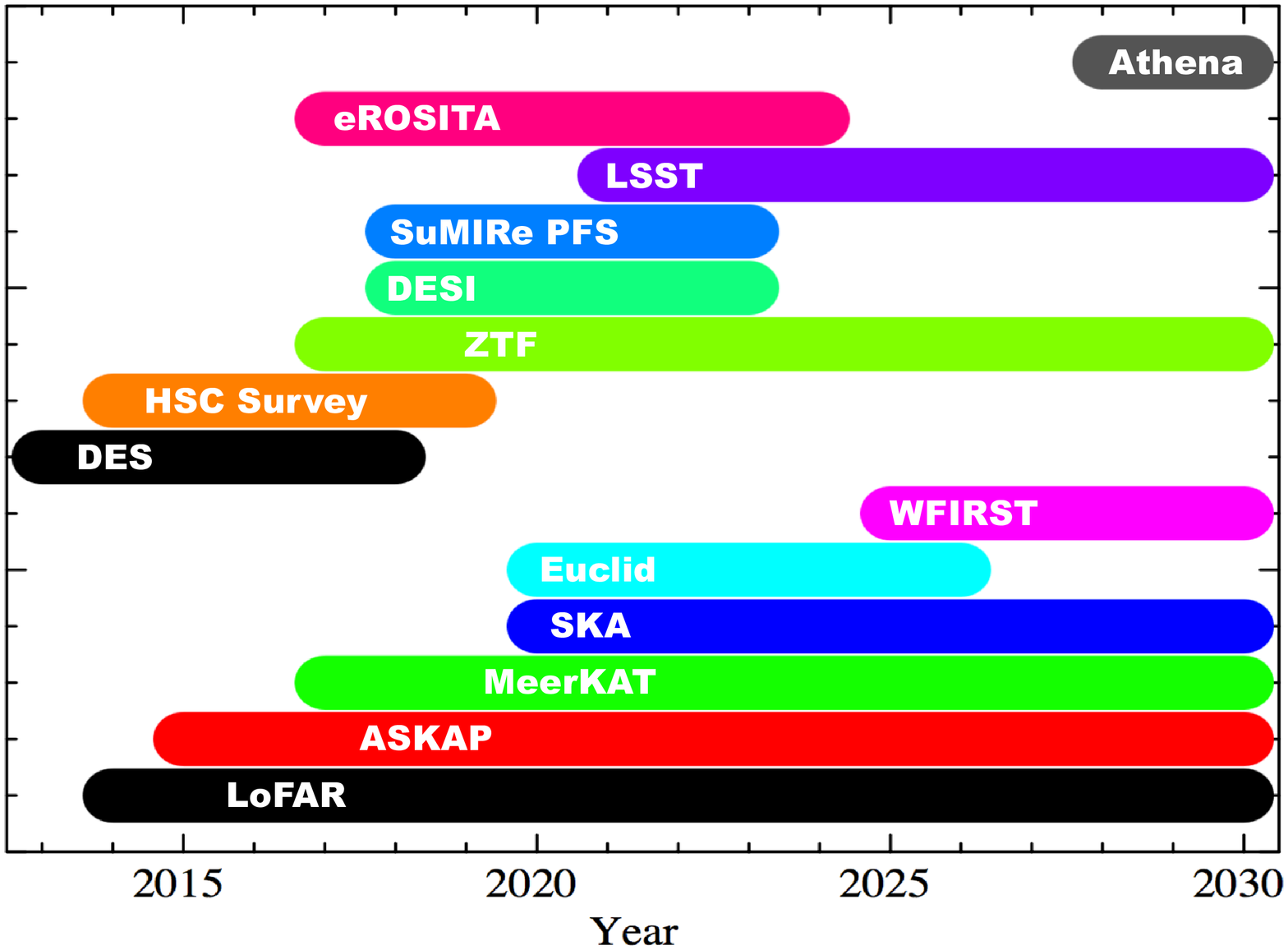}
\includegraphics[width=70mm,height=82mm,angle=-90]{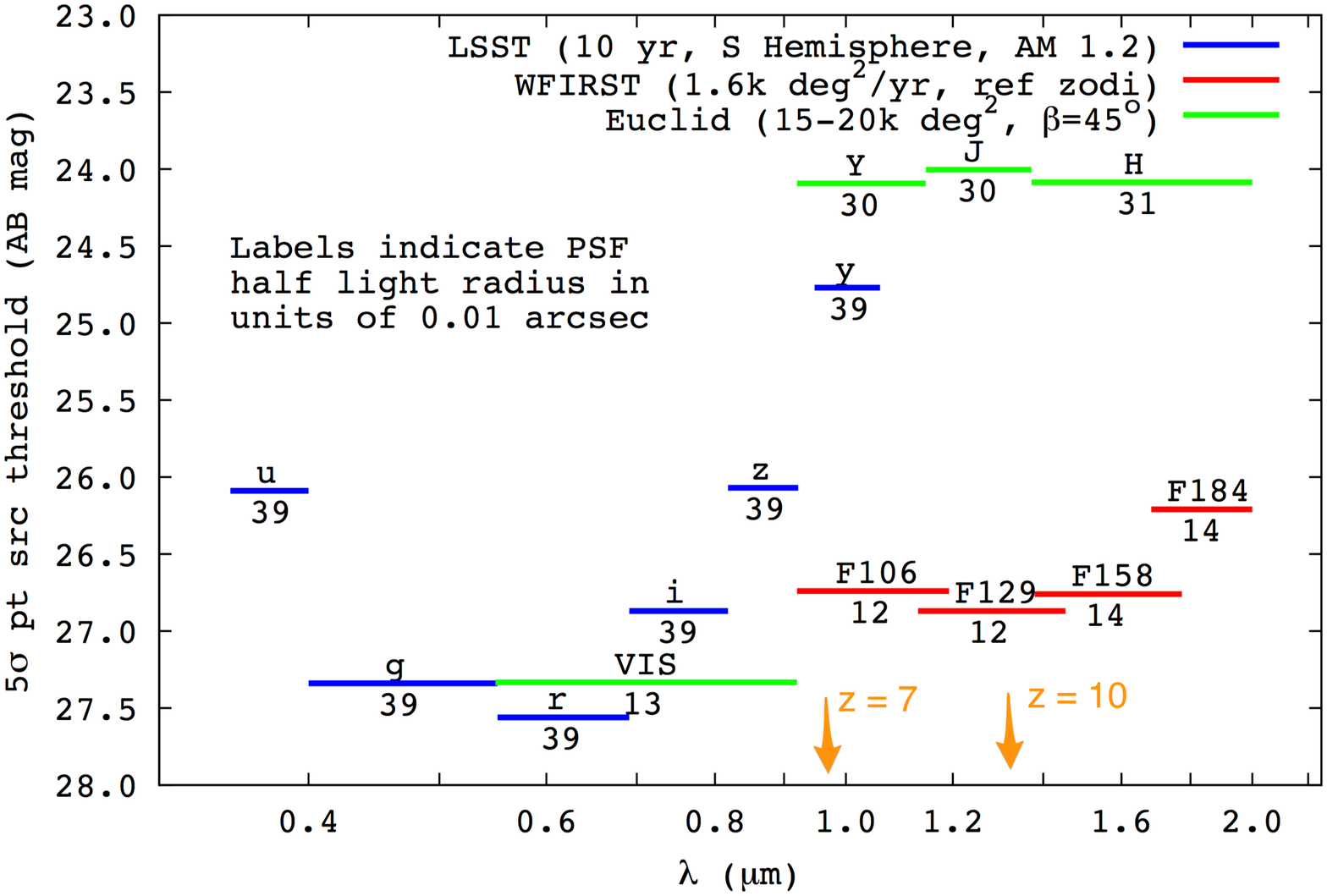}
\caption{{\it (Top)} Gantt chart showing expected operation dates for some 
selected future large survey projects. Dates are 
approximate and become increasingly uncertain further
into the future. Some listed projects will likely extend beyond 2030. 
For comparison, the next decade of \chandra\ and \xmm\ corresponds to 
\hbox{2016--2026}.
{\it (Bottom)} Expected depths for the LSST (10~yr), \euclid, and 
\wfirst\ surveys. Labels above each bar show photometric bands, 
and labels below each bar show the PSF 50\% encircled-energy 
radius in units of 0.01~arcsec. The orange arrows near the bottom
of the panel show the location of Ly$\alpha$ at $z=7$ and $z=10$. 
Note the generally good match between LSST and \wfirst\ 
depths. Adapted from Spergel et~al.\ (2013), where many 
further details are provided.}
\label{comingfacilities}
\end{figure}

Fig.~\ref{comingfacilities} is helpful to review when considering \xray\ survey 
prospects for the next decade of \chandra\ and \xmm. It shows there will
be a flood of very wide field multiwavelength photometric and optical/near-infrared
spectroscopic data becoming available. While some of these projects
have ``first light'' dates well into the future that are thus uncertain, 
others are more definite. For example, the Large Synoptic Survey Telescope (LSST) is 
now under active construction, and its deep-wide $ugrizy$ main survey should 
discover $\approx 20$ million AGNs including $\simgt 400,000$ from \hbox{$z=4$--7.5}. 
Combining LSST data with data from \euclid\ and \wfirst\ (see the bottom panel 
of Fig.~\ref{comingfacilities}), it should be possible 
to discover many more AGNs including 
$\simgt 1000$ at $z>7$ and $\simgt 20$ at $z>10$
(e.g., Spergel et~al.\ 2013). Thus, over the 
next decade, the demographic details of the unobscured 
and moderately obscured AGN population should become well understood 
out to very high redshifts. 

X-ray studies can vitally complement this work by revealing the demographics
of highly obscured AGNs out to high redshifts, and one can imagine extremely
powerful high-redshift investigations that combine thousands of archival \chandra\ 
and \xmm\ observations with the sensitive and very wide field imaging from, e.g.,
the Dark Energy Survey (DES), Hyper Suprime-Cam (HSC), 
LSST, \euclid, and \wfirst\ (see Fig.~\ref{comingfacilities}). Hopefully, 
\erosita\ will also contribute key \xray\ data to this endeavor. In 
these investigations, AGNs detected in the archival \xray\ data would 
be identified as high-redshift candidates based on their superb 
optical/infrared photometric data using primarily the Ly$\alpha$ 
forest (see the bottom panel of Fig.~\ref{comingfacilities}).
The only new observational cost would be to perform 
optical/near-infrared spectroscopic follow-up of these candidates. Such 
follow-up could be partly obtained with next-generation wide-field 
spectroscopic surveys; e.g. the Dark Energy Spectroscopic Instrument 
(DESI) and the Prime Focus Spectrograph (PFS) of Subaru Measurement 
of Images and Redshifts (SuMIRe) as shown in 
Fig.~\ref{comingfacilities}. For the optically fainter AGNs, dedicated 
spectroscopy with future Extremely Large Telescopes could be 
obtained. 

\begin{figure}
\begin{center}
\includegraphics[width=82mm,height=70mm,angle=0]{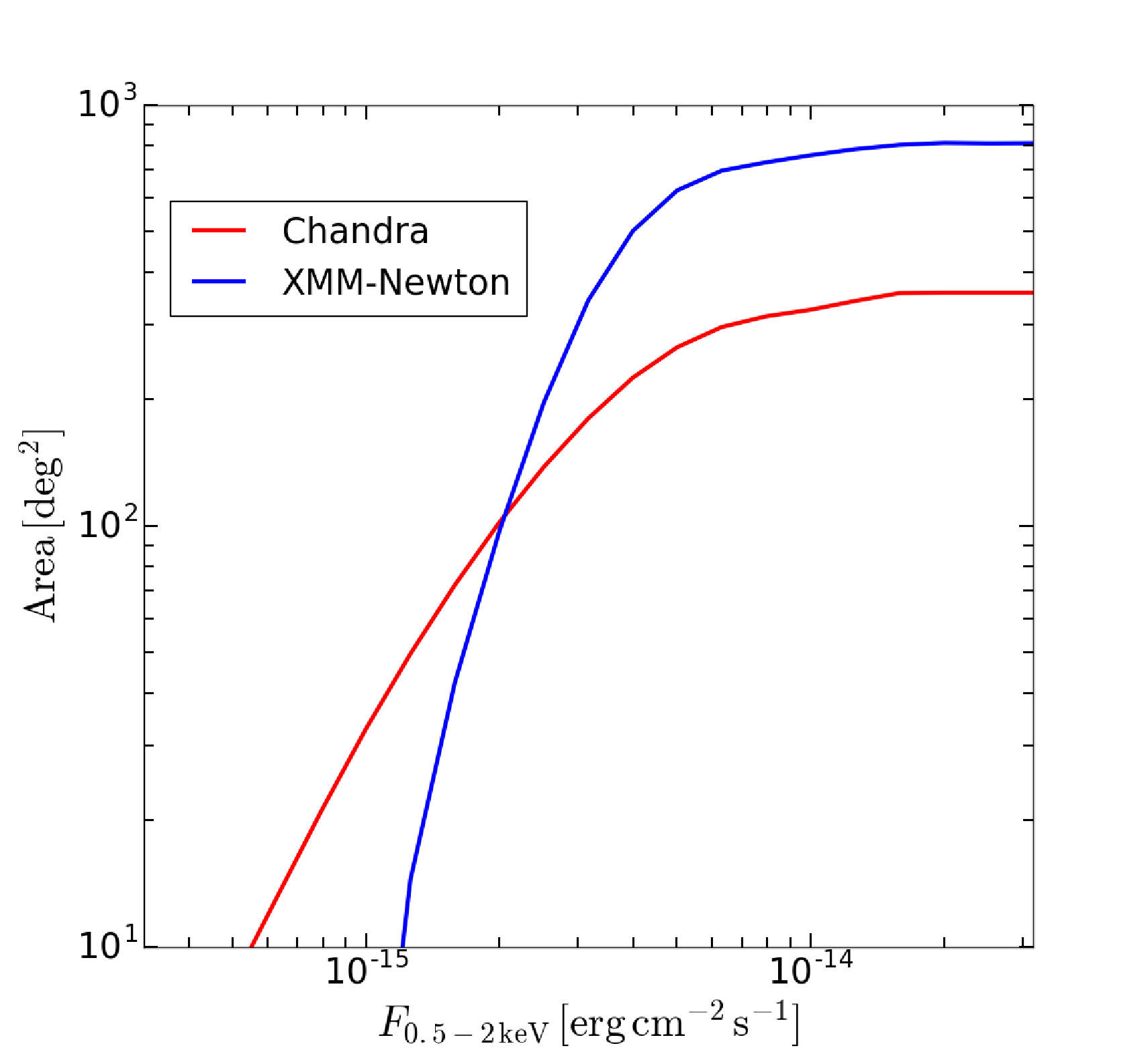}
\end{center}
\caption{Approximate solid-angle vs.\ depth plot for 
a massive 25~yr archival \chandra\ and \xmm\ survey that
could be performed in 2025. This has been derived by
scaling from the currently available archival data, 
assuming that the future distributions of exposure
times remain similar to the past ones. Only the deepest
60\% of \chandra\ and \xmm\ archival observations are 
considered; these are mostly \hbox{20--100}~ks exposures, 
but with a few exposures up to Ms levels. We have randomly 
removed $1/3$ of all observations to account for those 
not suitable for extragalactic surveys work 
(e.g., pointings in the Galactic plane, of bright
and extended foreground galaxies, or of bright 
foreground galaxy clusters). We have also approximately
accounted for the expected sensitivity loss over time
of \chandra\ and \xmm.}
\label{massivearchive}
\end{figure}

In 2025, one could plausibly aim to perform such a survey 
over $\approx 1200$~deg$^{2}$ utilizing the deepest \hbox{$\approx 60$\%}  
of \chandra\ and \xmm\ archival observations (see Fig.~\ref{massivearchive}).  
These would mostly be \hbox{20--100}~ks exposures, but 
with a few exposures up to Ms levels.  
Such a massive archival survey, using 
$\approx 25$~yr of data from both \chandra\ and \xmm, should 
detect \hbox{500--1100} AGNs 
at \hbox{$z=4$--6} and \hbox{10--100} at \hbox{$z=6$--8}, 
considering both obscured and unobscured systems.  
The detected AGNs will be relatively luminous with 
\hbox{2--10~keV} luminosities of \hbox{$L_{\rm X}=10^{44}$--$10^{45}$~erg~s$^{-1}$} 
at \hbox{$z=4$--6} and $L_{\rm X}=10^{44.5}$--$10^{45.5}$~erg~s$^{-1}$ at 
\hbox{$z=6$--8} (quoted luminosity ranges cover $\approx 90$\% of the
expected detections), providing quality constraints on the 
\xray\ luminosity function and the \xray\ obscured fraction 
($f_{\rm obsc}$) in this regime. The constraints on the 
$f_{\rm obsc}$-$L_{\rm X}$ and $f_{\rm obsc}$-$z$ relations 
require considerable improvements at the highest 
luminosities and redshifts. 
At lower AGN luminosities of $L_{\rm X}<10^{44}$~erg~s$^{-1}$, 
setting detailed constraints on the \hbox{$z=4$--8} \xray\ 
luminosity function will require deeper \xray\ observations 
over large areas with, e.g., \athena\ and \xrs. 

The massive archival survey described above, by dint of its
large solid-angle coverage, will automatically sample a wide
range of high-redshift cosmic environments. However, even 
it will tend to undersample rare regions such as the most
overdense structures in the early universe, where simulations 
predict that high-rate SMBH accretion can be sustained by 
protogalaxy merger episodes or continuous accretion of cold 
gas (e.g., Li et~al.\ 2007; Di~Matteo et~al.\ 2012; 
Costa et~al.\ 2014). The sampling of such regions can be 
improved via targeted deep \chandra\ and \xmm\ observations.
One such \chandra\ program is ongoing to determine the AGN 
content in the overdensity around the $z=6.28$ quasar 
SDSS~J1030+0524, and additional impressive overdensities at 
$z>4$ will surely be identified by the wide-field surveys shown 
in Fig.~\ref{comingfacilities}.  

\subsection{X-ray stacking of JWST galaxy samples}

The next decade will also bring substantially improved \xray\
stacking studies that constrain further the average amount of
SMBH accretion out to the highest redshifts. These will combine
the deepest \xray\ surveys with, e.g., high-redshift galaxy samples 
found in \jwst\ observations. 

The extremely deep photometric and spectroscopic data 
provided by \jwst\ will greatly improve the quality
of the galaxy samples being stacked, allowing better redshift
identifications and improved removal of low-redshift interlopers. 
Even more importantly, the redshift range of the galaxy samples
will be extended up to $z\approx 15$ with good source statistics
(compare with Fig.~\ref{stacking}), breaking into the critical 
redshift range where stacking constraints on the faint end of the \xray\ 
luminosity function will directly probe SMBH seeding mechanisms. 
The seeds could either be the relics of Pop~III stars, formed with 
a top-heavy initial mass function and ending up in black holes 
with masses of $\sim 100$~$M_{\odot}$, or $10^{4-6} M_{\odot}$ 
black holes formed from the collapse of primordial gas clouds
(e.g., Johnson \& Haardt 2016; Volonteri et~al.\ 2016). 
Such stacking investigations will lay critical groundwork for \xrs.

\section{X-ray spectroscopy and high-redshift AGN physics}

\subsection{Obscured protoquasars and host-galaxy feedback}

\begin{figure}
\includegraphics[width=80mm,height=70mm,angle=0]{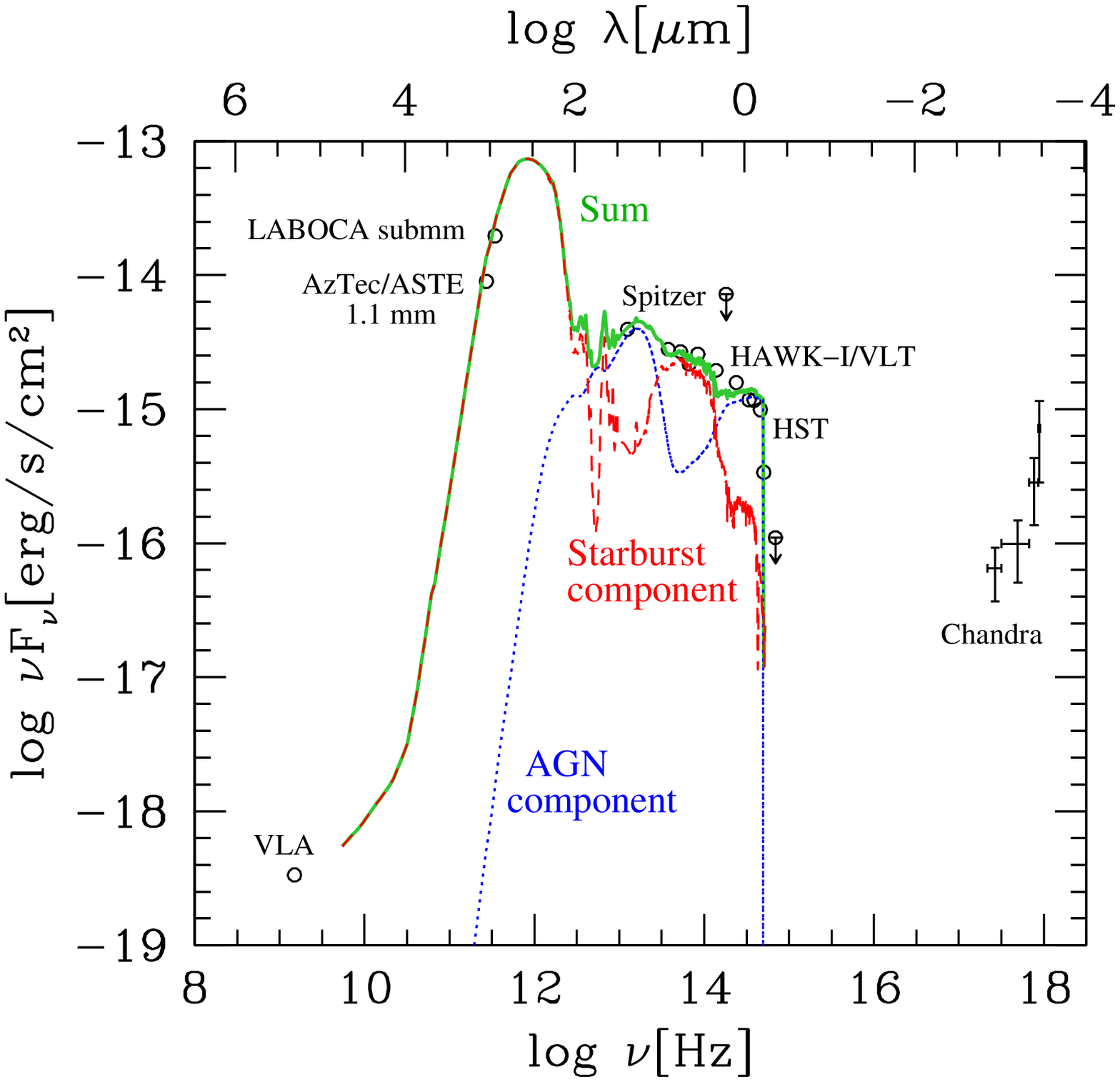}
\includegraphics[width=80mm,height=70mm,angle=0]{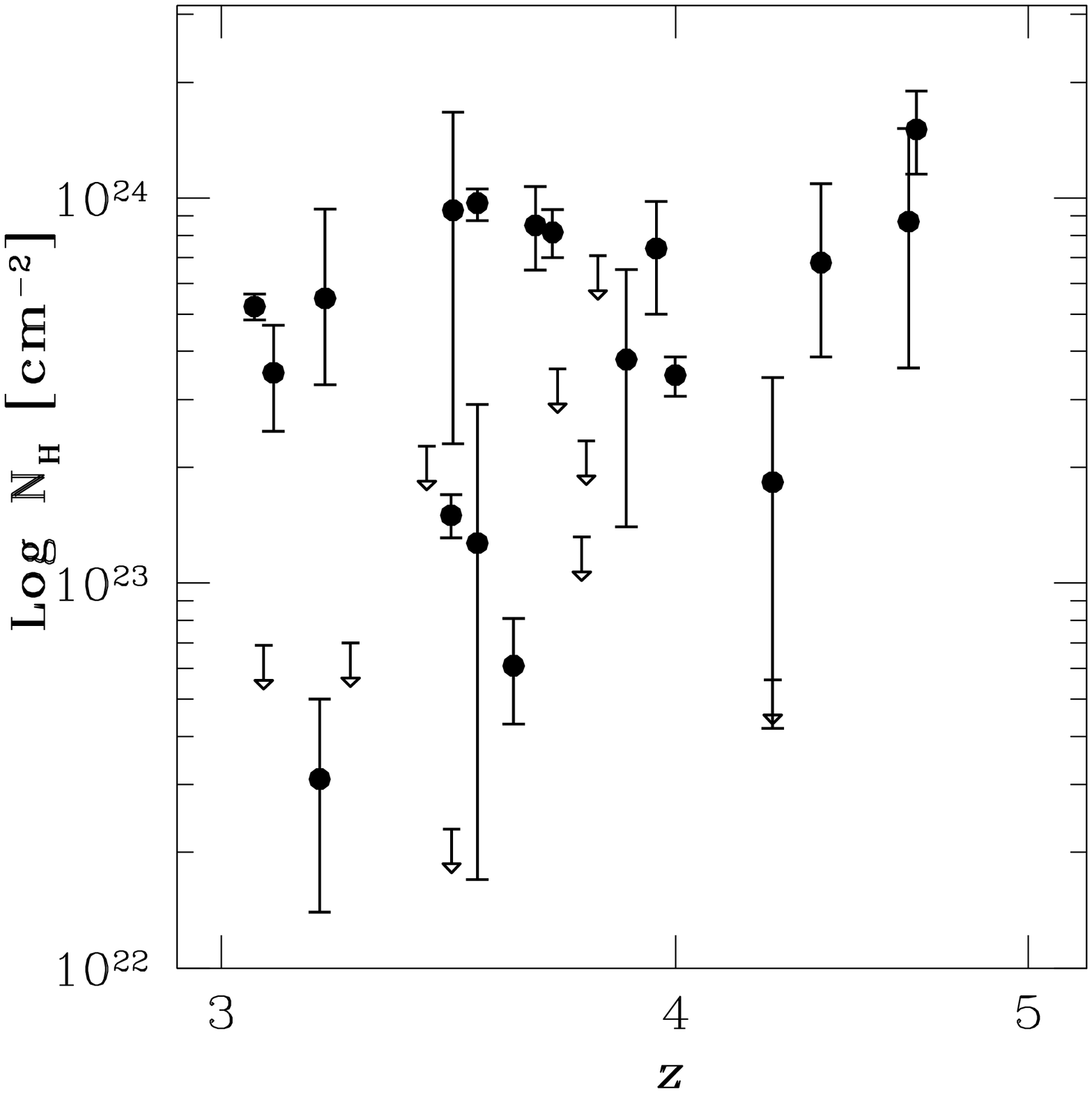}
\caption{{\it (Top)} Observed-frame radio-to-X-ray SED of a spectroscopically 
confirmed Compton-thick AGN at \hbox{$z=4.76$} in the \hbox{CDF-S}. The 
SED has been decomposed into an AGN component (blue dotted line) 
and a starburst-galaxy component (red dashed line); the green solid line 
is the sum of the two. Adapted from Gilli et~al.\ (2011).
{\it (Bottom)} Best-fitting column density vs.\ redshift for high-redshift \hbox{CDF-S}  
AGNs (upper limits are shown as downward pointing arrows). More than half
of these \xray\ selected AGNs are highly obscured with $N_{\rm H}>10^{23}$~cm$^{-2}$. 
Adapted from Vito et~al.\ (2013, 2014).}
\label{obscuredagn}
\end{figure}

A commonly considered picture for the formation of the first 
SMBHs at high redshifts involves the growth via accretion of lower 
mass black-hole seeds in gas-rich and frequently merging protogalaxies 
(e.g., Li et~al.\ 2007). As these seeds grow into SMBHs, their 
moderate-luminosity emission is initially obscured 
(``obscured protoquasars'') until one of the SMBHs grows to a mass
where it can provide effective feedback, likely via a radiation-driven 
wind. At this point, cold gas is expelled from the host galaxy by the 
wind, revealing a luminous unobscured AGN. Eventually, the feedback may 
expel sufficient gas to limit further SMBH growth and host-galaxy
star formation. 

X-ray spectroscopic investigations can provide insights into 
and constraints upon this picture. First, \xray\ spectroscopic 
studies of moderate-luminosity \hbox{$z=3$--5} AGNs in the 
deepest \xray\ surveys commonly ($\simgt 50$\%) show heavy 
obscuration with $N_{\rm H}>10^{23}$~cm$^{-2}$ 
(see Fig.~\ref{obscuredagn}), consistent with expectations for 
obscured protoquasars. Additional deep surveys with \chandra\ 
and \xmm, obtaining sufficient photon statistics for 
robust source spectral characterization, can improve constraints 
upon this fraction at \hbox{$z=3$--5} and its luminosity dependence. 

Furthermore, as observed at lower redshift, some
luminous high-redshift quasars have shown relativistic outflows
revealed primarily via iron~K absorption lines. These outflows 
likely possess sufficient mass-outflow rates and kinetic powers 
to expel large masses of cold gas from host galaxies, as needed
in the picture above. The prototypical example of such a system
is the gravitationally lensed quasar APM~08279+5255 at $z=3.91$, 
where a relativistic wind with velocities of \hbox{0.1--0.4$c$}, 
or larger, is detected (Chartas et~al.\ 2002, 2009). Modeling 
suggests a mass-outflow rate of 
\hbox{10--60~$M_\odot$~yr$^{-1}$} and 
a kinetic luminosity of $\approx 10^{46}$~erg~s$^{-1}$. 
Such outflows could be present, but undetected, in many other
high-redshift quasars---for APM~08279+5255, the 
lensing provides uniquely high \xray\ spectral quality. 
Deep \xray\ spectroscopy of a few additional carefully selected
quasars at \hbox{$z\approx 4$--6} could search for similar
wind-feedback signatures. 

\subsection{Measuring early SMBH accretion}

The very observation of \hbox{$10^8$--$10^{10}$~$M_{\odot}$} SMBHs 
in quasars at \hbox{$z=4$--7}, less than \hbox{1--2~Gyr} after 
the Big Bang, sets tight constraints on models for the formation 
and growth of early SMBHs. In order to explain the large SMBH 
masses observed up to \hbox{$z=6$--7}, one would need almost 
uninterrupted, Eddington-limited, accretion of gas from 
$z\approx 20$ to $z\approx 6$ with fairly low radiative efficiency.
Given the challenge arising from the first quasars, it is of natural 
interest to determine the accretion properties of high-redshift
AGNs, searching for, e.g., Eddington-limited accretion.  

X-ray observations can provide unique insights about the inner 
accretion disk and its corona out to high redshift. For 
example, the ratio between the \xray\ 
and optical/UV luminosities measures the relative 
importance of the accretion-disk corona vs.\ the disk itself 
(see \S4 of Brandt \& Alexander 2015 for a recent review). 
This ratio is usually parameterized by \aox\ which is the 
slope of a nominal power law joining rest-frame 2500~\AA\ and 2~keV  
[i.e., $\alpha_{\rm ox}=0.38 \log(L_{\rm 2~keV}/L_{2500~\mathring{\rm{A}}})$]. 
A significant correlation between \aox\ and the UV luminosity 
($L_{2500~\mathring{\rm{A}}}$) has been known since early studies, 
and this correlation has now been measured precisely with large 
samples of optically (e.g., Steffen et~al.\ 2006; Just et~al.\ 2007) 
and \xray\ (e.g., Lusso et~al.\ 2010) selected AGNs. 
Luminous quasars have spectral energy distributions (SEDs) with 
steeper (i.e., more negative) \aox\ indices indicating the 
dominance of the disk emission with respect to the \xray\ luminosity 
produced in the corona. It is thus common also to utilize
$\Delta\alpha_{\rm ox}=\alpha_{\rm ox}({\rm Observed})-\alpha_{\rm ox}(L_{2500~\mathring{\rm{A}}})$, 
which usefully quantifies the observed \xray\ luminosity relative 
to that expected from the \hbox{\aox-$L_{2500~\mathring{\rm{A}}}$} relation. 
Most recent studies find no significant evolution of 
\aox\ or \daox\ with redshift out to \hbox{$z\approx 5$--6}, beyond 
which the source statistics become too limited for robust constraints
(e.g., Just et~al.\ 2007; Lusso et~al.\ 2010; but see 
Kelly et~al.\ 2007). The tightest limits upon the \hbox{\aox-$z$} relation 
imply that variations of the typical $L_{\rm X}/L_{\rm UV}$ ratio with 
redshift are less than 30\%. This lack of strong \aox\ and \daox\ 
evolution is broadly consistent with the similar lack of 
evolution in other quasar properties generally: e.g., 
infrared continuum emission (e.g., Jiang et~al.\ 2006, 2010), 
emission-line strengths (e.g., De~Rosa et~al.\ 2011; Fan 2012), and 
\xray\ variability (e.g., Shemmer et~al.\ 2014; Yang et~al.\ 2016). 

The photon index of the hard \xray\ power-law continuum ($\Gamma$), 
which also measures the coupling between disk emission and the 
overlying hot corona, seems empirically to be a more robust 
tracer of the accretion rate than \aox. After the pioneering 
work of Shemmer et~al.\ (2006), the relation between $\Gamma$ and 
the Eddington ratio ($\lambda_{\rm Edd}=L/L_{\rm Edd}$) has been 
established over a range of redshifts for sizable samples of sources; 
steeper slopes correspond to higher implied $\lambda_{\rm Edd}$ 
(e.g., Shemmer et~al.\ 2008; Risaliti et~al.\ 2009; 
Brightman et~al.\ 2013; Fanali et~al.\ 2013). The current
studies of \hbox{$z\approx 4$--5.5} radio-quiet quasars generally 
do not indicate exceptional $\Gamma$ or $\lambda_{\rm Edd}$ values 
(e.g., Shemmer et~al.\ 2005; Vignali et~al.\ 2005). However, 
there are hints of potentially steep $\Gamma$ values for three 
quasars at $z=6.28$, $z=6.30$, and $z=7.08$, one of which is 
debated (Farrah et~al.\ 2004; Ai et~al.\ 2016; 
Moretti et~al.\ 2014 vs.\ Page et~al.\ 2014). 

All current \xray\ spectral studies of $z>4$ quasars 
suffer from limited photon statistics, and major observational
investments with \chandra\ and \xmm\ could improve this 
situation allowing more general and reliable 
conclusions about \xray\ continuum shapes to be drawn. This 
is especially true at the highest redshifts of $z>6$. As 
high-redshift quasars are faint and thus subject to 
significant damage when satellite background flaring occurs, 
any further flaring accommodations would be helpful to ensure 
that such approved observations actually achieve their full 
proposed exposures. In the more distant future, as luminous 
quasars at \hbox{$z=7$--10} are discovered by the combination of 
DES, HSC, LSST, \euclid, and \wfirst\ (see \S2.2 for details), 
these will be critical targets for pushing \xray\ spectral 
constraints deep into the reionization era. 

\subsection{Extreme AGN sub-populations at high redshift}

Extreme AGN sub-populations can often ultimately be used
as tools to teach us about the broader population more generally. 
Such extreme objects can reveal accretion phenomena that are 
generally applicable but are difficult to identify when more  
subtly expressed in the overall population. 
There are several extreme AGN sub-populations
that have been identified at $z>4$ (and often also at lower
redshifts), including 
weak-line quasars (e.g., Diamond-Stanic et~al.\ 2009; Luo et~al.\ 2015), 
hot-dust poor quasars (e.g., Jiang et~al.\ 2006, 2010), and
highly radio-loud quasars (e.g., Wu et~al.\ 2013; Ghisellini et~al.\ 2015), 
and \xray\ studies of these can provide insights about their
nature. 

X-ray observations of a large sample of weak-line 
quasars support a model where these quasars possess thick
inner accretion disks, perhaps due to high $\lambda_{\rm Edd}$
(Luo et~al.\ 2015). This model can explain, in a simple and 
unified manner, their weak lines and 
diverse \xray\ properties. Further
\xray\ spectroscopy of selected weak-line quasars can 
search for spectral signatures associated
with the thick inner disk. Finally, highly radio-loud quasars
at $z>4$ (including some blazars), which 
launch the most-powerful jets made by growing 
SMBHs in the early universe, appear $\approx 3$ times \xray\ 
brighter than their matched counterparts at lower redshifts
(Wu et~al.\ 2013). This may be due to a fractional IC/CMB 
contribution (inverse Compton scattering of the cosmic microwave 
background) to the jet-linked core \xray\ luminosity that grows rapidly 
with redshift. Further \chandra\ and \xmm\ observations/analyses
can establish definitively these high-redshift \xray\ enhancements 
and clarify their dependence upon redshift, allowing testing of 
the fractional IC/CMB model. 

\acknowledgements
We thank R. Gilli and C. Vignali for helpful feedback. 
We acknowledge support from 
\chandra\ \xray\ Center grants GO3-14100X and GO4-15130A 
as well as the V.M. Willaman Endowment. 

\newpage


\end{document}